\documentclass{JAC2003}

\usepackage{graphicx}
\usepackage{booktabs}

\begin{document}
\title{RETENTION OF ATTOSECOND PULSE STRUCTURE IN AN HHG SEEDED FEL AMPLIFIER}

\author{B. W. J. McNeil, SUPA, University of Strathclyde, Glasgow G4 0NG, UK,\\
N. R. Thompson and D. J. Dunning, ASTeC, Daresbury Laboratory, Warrington WA4 4AD, UK,\\
 B. Sheehy, Sheehy Scientific Consulting, New York, USA. }

\maketitle

\begin{abstract}
A model is presented which demonstrates that the attosecond pulse structure of a High Harmonic Generation (HHG) seed may be retained through to saturation in an FEL amplifier. At wavelengths of $\sim12$~nm, a train of attosecond pulses of widths $\sim300$~attoseconds with peak powers in excess of 1~GW are predicted from full 3D simulation. Methods for improving these results are discussed.\end{abstract}

\section{Introduction}
High-gain free-electron lasers (FELs) operating at short wavelengths have the potential to gather incredibly detailed information on how matter interacts and arranges itself at the atomic and molecular scales~\cite{attosecondscience}. If high brightness pulses of sufficiently short duration ($~10^{-15}-10^{-17}$s) and power can be generated, it may also be possible to capture ephemeral events, such as molecular bond formation, without temporal blurring. While HHG delivers trains of attosecond pulses, they currently lack the peak powers available from FELs.

Several techniques to achieve such pulse durations have been proposed (see e.g.\ \cite{saldin,zholentsNJP,emma} and references therein) where the widths of the pulses generated are of the order of the coherence length $l_{c}=\lambda_r/4\pi\rho$, where $\lambda_r$ is the resonant wavelength and $\rho$ is the fundamental FEL parameter~\cite{bnp}.

A different technique, that applies the concepts from mode-locked cavity lasers, has also been proposed~\cite{mlsase} that may generate a train of pulses with lengths less than $l_{c}$. In this technique, a series of spatiotemporal shifts are introduced between the radiation and the co-propagating electron bunch that define a set of axial radiation modes which couple via the FEL interaction. The spatiotemporal shifts are achieved by delaying the electron bunch using magnetic chicanes inserted between undulator modules. By introducing a modulation at the mode spacing, each mode develops sidebands that can overlap neighbouring modes, allowing mode-locking to occur. This is analogous with mode-locking in conventional cavity lasers.

The spectrum generated by the XUV mode-coupled SASE system of ~\cite{mlsase} has a modal structure with modespacing $\Delta\omega$ similar to the spacing of harmonics in a HHG source. Equivalently, the temporal structure of the two systems are a comb of attosecond pulses. By matching the spectral/temporal structures of a mode-coupled FEL amplifier to an HHG seed, it may therefore be possible to amplify the HHG seed to saturation while retaining its attosecond train structure. Furthermore, it should be possible to achieve this without the need for introducing a modulation at the mode spacing as the HHG attosecond seed pulses already have a well-defined phase relationship. In this paper it is demonstrated via 1D and 3D numerical simulations that this is the case.

\section{MODE GENERATION IN AN AMPLIFIER}
\begin{figure}
\includegraphics[width=80mm]{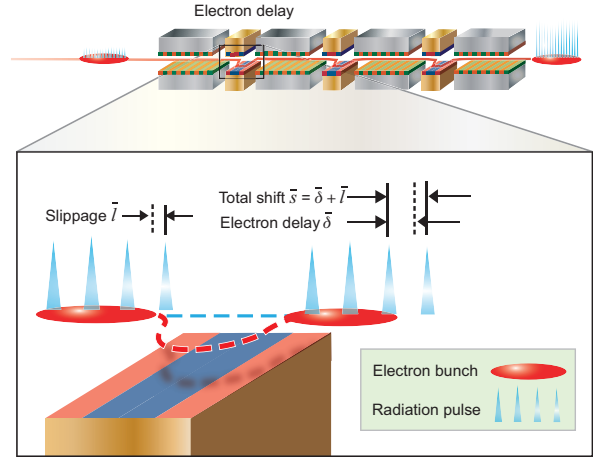}
\caption{\label{fig1}Schematic of the mode-coupled amplifier FEL interaction operating with an HHG seed. The inset shows detail of the electron delay.}
\end{figure}

The spatio-temporal shifts that generate the equally spaced modal structure in the amplifier spectrum are achieved by
periodically delaying the electron bunch using magnetic chicanes between undulator modules~\cite{mlsase}, as shown schematically in
Fig.~\ref{fig1}.  The discrete modal spectrum is similar to the axial modes of a conventional laser.
These modes are intrinsically coupled via their collective interaction with the electron bunch over regions of a coherence
length $l_{coh}\approx l_c$ where $l_c=\lambda_r/4\pi\rho$ is the FEL cooperation length.

The notation used here follows that of~\cite{mlsase}. The total slippage of the radiation with respect to the electron bunch per module in units of the scaled electron bunch co-ordinate $\bar{z}_{1}$~\cite{mlsase} is $\bar{s}=\bar{l}+\bar{\delta}$, where $\bar{l}$ is the slippage occurring in the undulator and  $\bar{\delta}$ is the slippage within the chicane. The  scaled resonant FEL frequency is $\bar{\omega}=(\omega-\omega_{r})/2\rho\omega_{r}$ where $\omega_{r}$ is the resonant FEL frequency. The slippage enhancement factor is defined as $S_{e}=\bar{s}/\bar{l}$. In~\cite{mlsase}, a solution is obtained for the one dimensional wave equation describing the field evolution of a small, constant electron source term for a series of $N$ undulator-chicane modules.
For $S_{e} > 1$ the spectrum takes the form of a sinc-function envelope (the single undulator module spectrum) modulated by a ''frequency comb'' centered at the scaled resonant frequency $\bar{\omega}$, with mode separation $\Delta\bar{\omega}=2\pi/\bar{s}$ corresponding to a $\Delta\omega=2\pi/T_{s}$ where $T_{s}=s/c=\bar{s}l_{c}/c$ is the time taken for radiation to travel the slippage length. The number of modes under the central peak of the spectrum is shown to be $N_{0}=2S_{e}-1$

\section{Amplification of an HHG seed}

The temporal profile of the HHG seed is a comb of attosecond pulses separated by half the wavelength of the drive laser, $\lambda_{d}$. The spectral and temporal structure of the radiation generated by the undulator-chicane system is matched to that of the HHG seed by setting $\bar{s}=\bar{\lambda}_{d}/2$, where $\bar{\lambda}_{d}=\lambda_{d}/l_c$ is the scaled drive laser wavelength.

For a typical HHG drive laser wavelength of $\lambda_d$~=~805~nm (e.g.\ Ti:Sapphire) and an FEL parameter of $\rho=2\times 10^{-3}$ similar to that for an FEL operating in the XUV, then for undulator modules of 8 periods $\bar{l}=0.201$. The value of the scaled slippage generated by the chicanes is chosen to be $\bar{\delta}=0.616$ so that $\bar{s}=\bar{\lambda}_{d}/2$, matching the undulator-chicane system to the spectral and temporal structure HHG seed. This gives a slippage enhancement factor of $S_{e}=4.0625$.

\begin{figure}[htb]
   \centering
   \includegraphics[width=80mm]{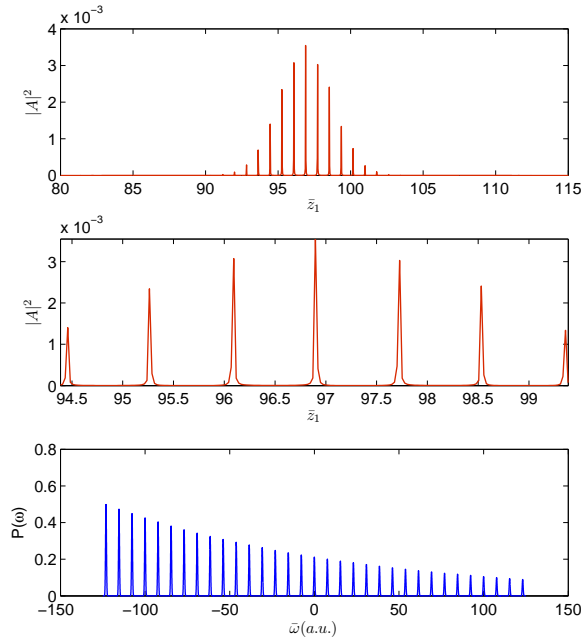}
   \caption{Scaled longitudinal intensity profile (top and middle) and scaled spectral power distribution for the HHG seed (bottom).}
   \label{seed}
\end{figure}

The HHG seed was modelled in a similar way to that described in~\cite{njpxuv} with the resonant wavelength chosen to be the 65th harmonic of the drive laser i.e.\ $\lambda_r=\lambda_d/65$. The Nyquist theorem limits the frequencies that can be modelled in the averaged code~\cite{njpxuv} to harmonics 33 to 97. The temporal and modal structure of the seed is shown in Fig.~\ref{seed}.

The system is modelled using a 1D code similar to that described in~\cite{felo}. The HHG seed is injected into the amplifying undulator-chicane modules with a cold electron beam. Note that no energy modulation of the electron beam was used. This is different to the mode-locked SASE case of~\cite{mlsase} which starts up from noise and requires a beam energy modulation at the mode spacing $\Delta\bar{\omega}$ to couple the modes and generate locking between each mode. Here, the HHG seed has modes which already have a well-defined phase relationship. It will be seen that while this phase relationship between modes is not significantly altered during amplification (the temporal pulse train structure is retained) it appears that the lack of modulation at the mode spacing $\Delta\bar{\omega}$ does not allow the modes to `lock' and the number of modes receiving significant amplification is reduced. This leads to a broadening of the individual pulses and appears in agreement with the scaling observed in~\cite{mlsase} for the pulse width $\tau_p \propto 1/\sqrt{N_0}$ where $N_0$ is the number of amplified modes.

\begin{figure}[htb]
   \centering
   \includegraphics[width=80mm]{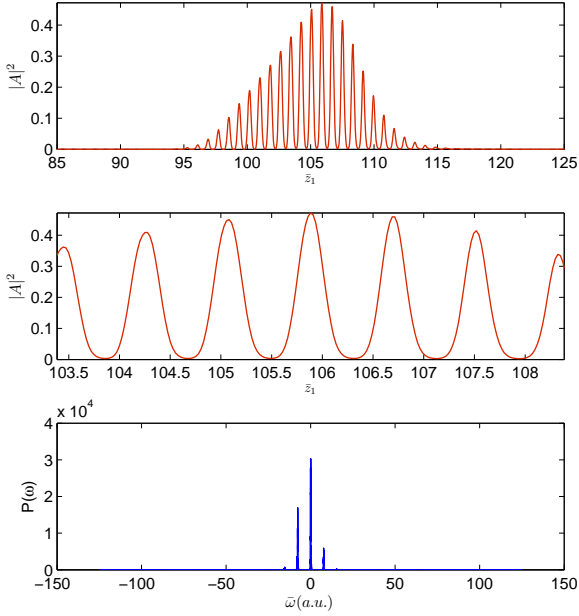}
   \caption{Scaled longitudinal intensity profile (top and middle) and scaled spectral power distribution of the amplified HHG radiation for $S_e=4.0625$, after 22 undulator-chicane modules.}
   \label{Se4_pMA_0}
\end{figure}

After 18 undulator-chicane modules the amplified scaled radiation power and spectrum is as shown in Fig.~\ref{Se4_pMA_0}. It is seen that the HHG seed is amplified and the comb of temporal spikes is retained, but with increased pulse widths as discussed. This number of modules corresponds to a FEL interaction length of $22\times\bar{l}\approx 4.4$. This is just prior to saturation where scaled powers of $|A|^2\approx 1$ are achieved. However, the temporal structure of the pulses begin to break up at saturation. (Note that the dispersive strength of the chicanes $D=R_{56}/2l_c\approx 10\bar{\delta}/6$ reduces the gain length as discussed in~\cite{mlsase}).

\section{Optimising the system}
In ~\cite{mlsase} it is demonstrated that the pulse widths in mode-locked SASE simulations are proportional to $1/\sqrt{N_{0}}$, this is in agreement with an analogous relation from mode-locked conventional cavity lasers.

In order to attain narrower amplified HHG pulses, the number of modes under the central peak of the spectrum is increased by increasing the value of the slippage enhancement factor. To increase $S_{e}$ while continuing to satisfy the condition $\bar{s}=\bar{\lambda}_{d}/2$, the value of $\bar{l}$ must be decreased while $\bar{\delta}$ is increased to keep $\bar{s}$ constant. The system is simulated for a case with the slippage enhancement $S_{e}=8.125$, with the number of undulator periods per module equal four.

\begin{figure}[htb]
   \centering
   \includegraphics[width=80mm]{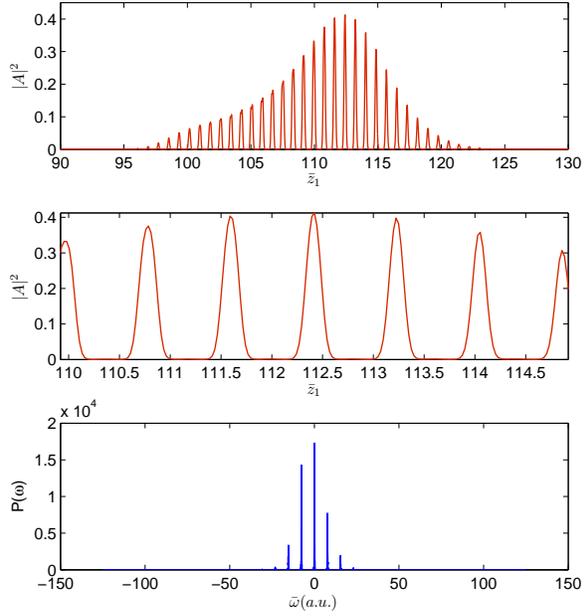}
   \caption{Scaled longitudinal intensity profile (top and middle) and scaled spectral power distribution of the amplified HHG radiation for $S_e=8.125$, after 34 undulator-chicane modules.}
   \label{Se8_pMA_0}
\end{figure}

The spectral power distributions for the two cases $S_{e}=4.0625$ and $S_{e}=8.125$ are shown in Figs.~\ref{Se4_pMA_0} and ~\ref{Se8_pMA_0} respectively. It is seen that the number of modes, $N_0$, under the central peak of the spectrum increases with $S_e$, narrowing the pulses. For $S_{e}=8.125$, 34 undulator-chicane modules are required to produce the same amplification of the 22 module $S_{e}=4.0625$ case.

For larger values of $S_{e}$, the broadening of the modal sinc-function envelope becomes restricted by the Nyquist limit.

\section{Energy Spread}

\begin{figure}[htb]
   \centering
   \includegraphics[width=80mm]{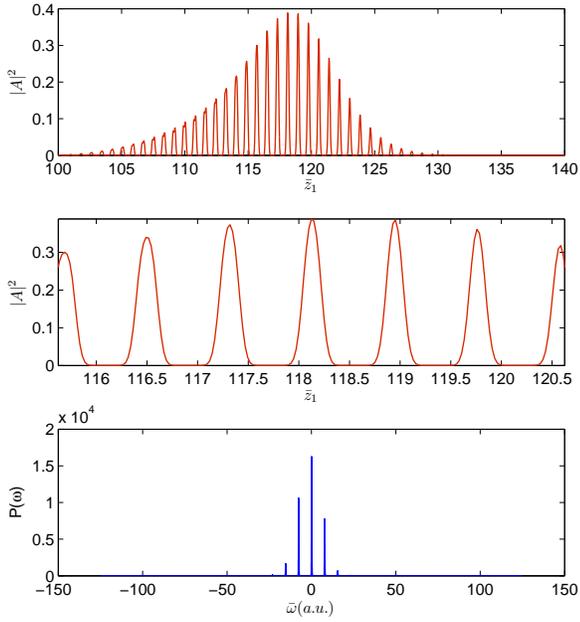}
   \caption{Scaled longitudinal intensity profile (top and middle) and scaled spectral power distribution of the amplified HHG radiation for $S_e=8.125$ and an initial energy spread $\sigma_{\gamma}/\gamma=0.02~\%$, after 42 undulator-chicane modules.}
   \label{Energy_spread}
\end{figure}

The effects of energy spread on the interaction have been investigated using the 1D model by introducing a relative gaussian energy spread of width $\sigma_{\gamma}/\gamma= \rho/10 = 0.02 \%$ for the case $S_{e}=8.125$.
The results close to saturation are plotted in Fig.~\ref{Energy_spread}, which is the equivalent of Fig.~\ref{Se8_pMA_0}. While the cold beam case took 34 undulator-chicane modules, the case here took 42 modules. Note that the temporal and spectral structure of the radiation power remains largely unaffected by the energy spread. Larger energy spreads have also been simulated with the main effect being a larger number of modules to saturation.

\section{Results of 3D Simulations}

The system has also been modelled using the three-dimensional code Genesis 1.3~\cite{genesis}. The main results of the 1D simulations are reproduced using the 3D code and show amplification of an HHG seed pulse retaining its pulse train structure.

\begin{figure}[htb]
   \centering
   \includegraphics[width=75mm]{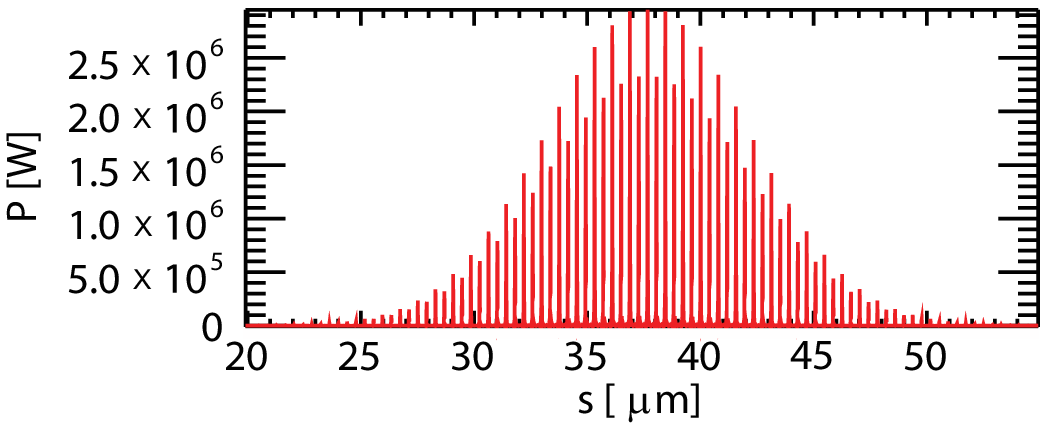}
   \includegraphics[width=75mm]{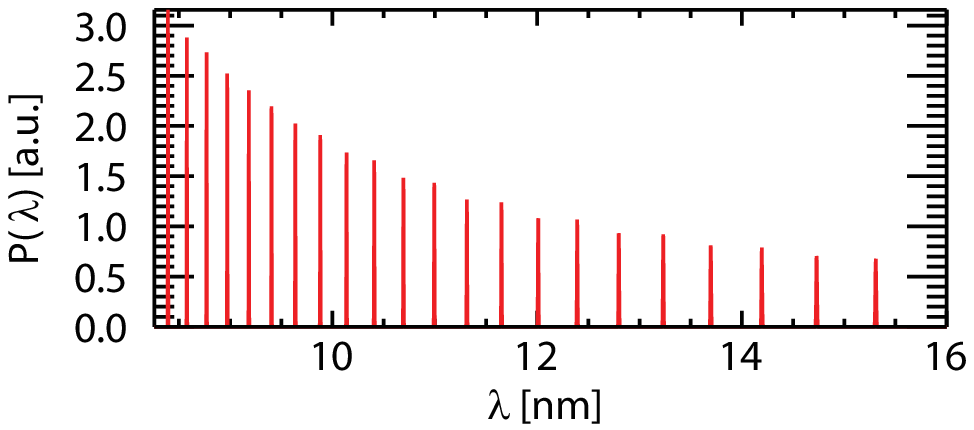}
   \caption{Longitudinal intensity profile (top) and spectral power distribution (bottom) of the HHG seed in 3D simulations.}
   \label{genesis_seed}
\end{figure}

Similar parameters as for the XUV mode-coupled FEL amplifier of~\cite{mlsase} were used (i.e. without electron beam modulation) including an energy spread of width $\sigma_{\gamma}/\gamma= 0.01 \%$. The seed power and spectrum show the attosecond pulse train structure and frequency comb, as seen in Fig.~\ref{genesis_seed}. The peak power of the seed is 3~MW with FWHM duration 33~fs. It was assumed that the wavelength of the Ti:Sapphire drive laser was $\lambda_d$~=~780~nm and the FEL was then tuned to be resonant with the 63rd harmonic of the drive laser with $\lambda_r$~=~12.4~nm.

\begin{figure}[htb]
   \centering
   \includegraphics[width=75mm]{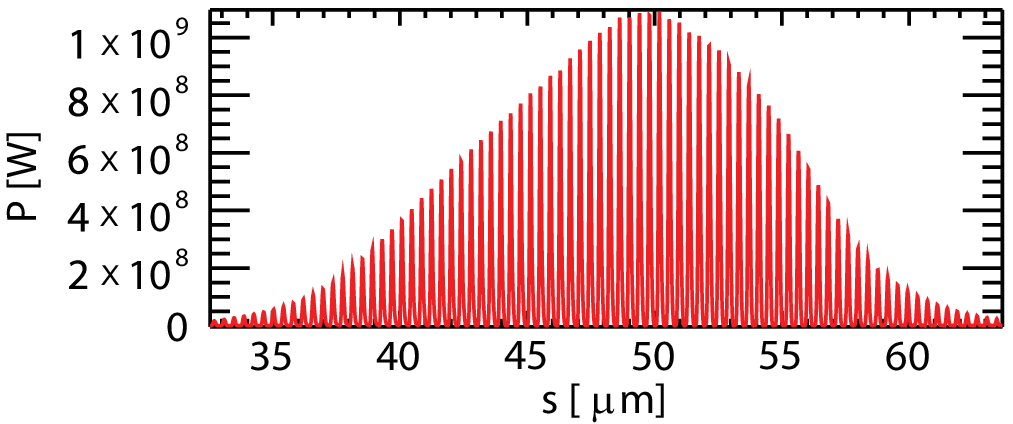}
   \vspace{1mm}
      \includegraphics[width=75mm]{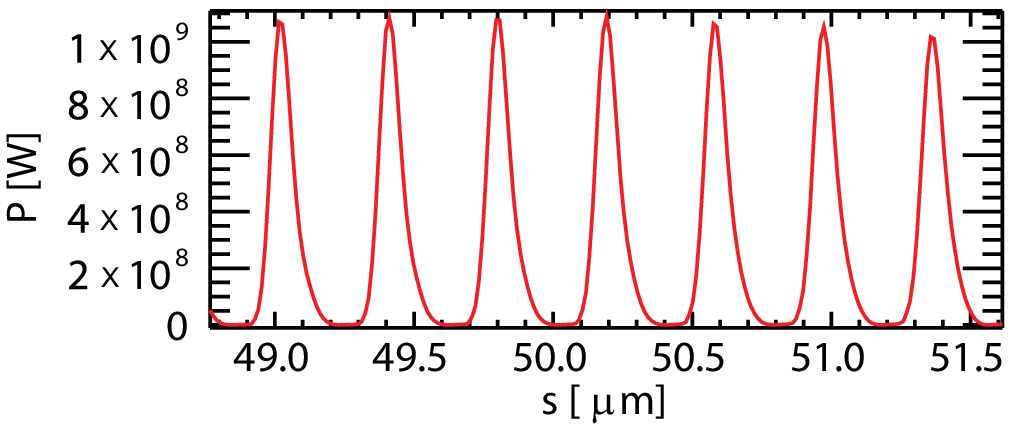}
         \includegraphics[width=75mm]{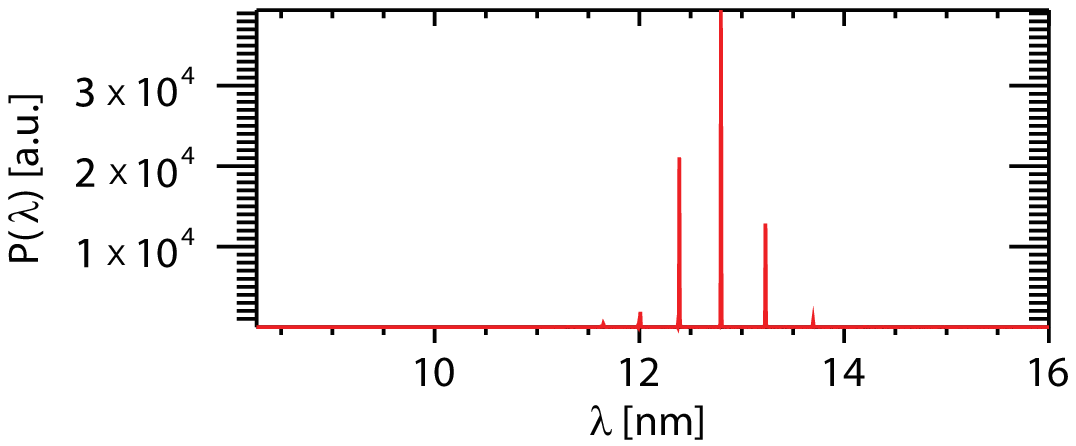}
   \caption{Longitudinal intensity profile (top and middle) and spectral power distribution (bottom) of the amplified HHG radiation in 3D simulations, with $S_e=8$. The agreement with the equivalent 1D simulations shown in Fig.~\ref{Energy_spread} is very good.}
   \label{genesis_amplified}
\end{figure}

Each undulator module had 4 periods of 3.1~cm and each chicane had a delay of 28 radiation wavelengths corresponding to a slippage enhancement $S_e=8$. These parameters for the undulator-chicane modules then match the mode spacing of the HHG seed. The amplified seed approaching saturation, after 48 modules, is plotted in Fig.~\ref{genesis_amplified} and the pulse temporal and spectral properties agree well with the equivalent one-dimensional $S_e=8.125$ case of Fig.~\ref{Energy_spread}. The FWHM  length of each pulse in the amplified train is $~$300~as. Further 3D simulations (not shown) with $S_e=4$ agreed well with the $S_e=4.0625$ case of Fig.~\ref{Se4_pMA_0}.

\section{Conclusions}
It is shown, using 1D and 3D simulations, that the attosecond structure of the HHG seed can be amplified to saturation in the mode-coupled amplifier FEL of ~\cite{mlsase}. It is stressed that no pre-conditioning of the electron beam (e.g. energy modulation) is required. The only requirement is for an undulator-chicane modular structure, the radiation spectrum of which matches an HHG seed source.

The introduction of an energy modulation to the system may well improve the amplification of the HHG to generate trains of narrower pulses by locking the modal structure. This will be investigated in future work.

Also to be investigated is the behaviour of the system when modelled in a non-averaged code, such as ~\cite{non-averaging}. Such codes have a larger Nyquist frequency and are capable of simulating a wider frequency range and so shorter pulse lengths.

\end{document}